\documentclass[aps,prl,twocolumn,superscriptaddress]{revtex4-1}

\usepackage[utf8]{inputenc}
\usepackage{amsmath}
\usepackage{graphicx}

\begin{document}

 
\title{Origin of nonclassical light emission \\
from defects in multi-layer hexagonal boron nitride}
 
 \author{Alexander~Bommer}
 \affiliation{Universit\"at des Saarlandes, Fachrichtung Physik, Campus E2.6, 66123 Saarbr\"ucken}
 
 \author{Christoph~Becher}
\affiliation{Universit\"at des Saarlandes, Fachrichtung Physik, Campus E2.6, 66123 Saarbr\"ucken} 

\date{\today}



\begin{abstract}In recent years, mono-layers and multi-layers of
hexagonal boron nitride (hBN) have been demonstrated as host
materials for localized atomic defects that can be used as
emitters for ultra-bright, non-classical light. The origin of
the emission, however, is still subject to debate. Based on
measurements of photon statistics, lifetime and polarization on
selected emitters we find that these atomic defects do not act
as pure single photon emitters. Our results strongly and
consistently indicate that each zero phonon line of individual
emitters is comprised of two independent electronic transitions.
These results give new insights into the nature of the observed
emission and hint at a double defect nature of emitters in
multi-layer hBN.
\end{abstract}


\maketitle

\section{Introduction}

Recently, two dimensional van der Waals materials have emerged
as promising platforms for optoelectronics
\cite{Srivastava2015,Palacios-Berraquero2016,Chakraborty2015},
candidates for future UV-LEDs \cite{Kenji2011,Watanabe2009} and
host materials for emitters of non-classical light
\cite{Tonndorf2015,Koperski2015,Tran2015,Tran2016a,Tran2016b,Martinez2016,Chejanovsky2016,Schell2016,Jungwirth2016,Exarhos2017}.
Especially atomic defects in hexagonal boron nitride have shown
to belong to the brightest emitters of non-classical light ever
reported. hBN is a semiconductor with a large band gap of around
6\,eV \cite{Cassabois2016}. Therefore, it is widely believed,
that at the origin of the emission are localized defects in the
host material that give rise to electronic transitions between
discrete energy levels within the band gap, as it is the case
for color centers in diamond \cite{Doherty2013,Neu2011}.
However, the exact nature of the defects still remains unclear
and is subject of ongoing experimental and theoretical
investigations \cite{Tawfik2017,Abdi2018,Sajid2018,Lopez2018}.
For application in quantum information one needs narrowband and
background free emission lines. The emitters selected in this
work fulfill these criteria and exhibit spectra consisting of an
asymmetric zero phonon line (ZPL) and a phonon side band
165\,meV red shifted from the ZPL. This energy shift corresponds
to a well-known phonon mode in hBN
\cite{Geick1966,Reich2005,Nemanich1981}. The asymmetry of the
ZPL is commonly attributed to phonon interaction and the ZPL
wavelengths have been shown to spread across a range from
500-800\,nm \cite{Dietrich2018}, which is attributed to strain
inside the host crystal \cite{Tran2016a,Grosso2017}. Independent
of the emission wavelength, the ZPL is assumed to consist of a
single, linearly polarized dipole transition giving rise to
single photon emission. In this publication, on the contrary, we
provide strong evidence for the presence of two independent
emitters in each defect and show that the second line causing
the asymmetry of the ZPL indeed is a second electronic
transition. By carefully evaluating photon correlation
measurements we see that we only are able to fully reproduce
our data by using an extended g$^{(2)}$-function, that takes
into account two independent transitions. We gain full access to
the parameters of the g$^{(2)}$-function via independently
measuring the spectra and the excitation power dependent photon
emission rates of the corresponding emitters. We further confirm
the existence of double defects via measuring polarization
dependent spectra and performing time correlated single photon
counting (TCSPC) measurements.

\section{Investigation of emission from point defects in hBN}
We spectroscopically investigate micrometer sized multi-layer
flakes of hexagonal boron nitride in a home built laser scanning
confocal microscope under continuous wave excitation at
$\lambda$ = 532\,nm. The commercially available flakes
(\textit{Graphene Supermarket}) are diluted in a solution
(50\,\% water, 50\,\% ethanol) with a concentration of
5.5\,mg/ml and put in an ultrasonic bath to break up
agglomerates. The solution is drop cast (5-10\,$\mu$l) onto a
silicon wafer with an iridium layer for enhanced photon
collection efficiency. The substrate is heated on a hotplate to
70$^{\circ}$C to evaporate the liquid. After drop casting,
individual flakes can be addressed in the confocal microscope.

\begin{figure}[h!]
	\centering
		\includegraphics[width=0.5\textwidth]{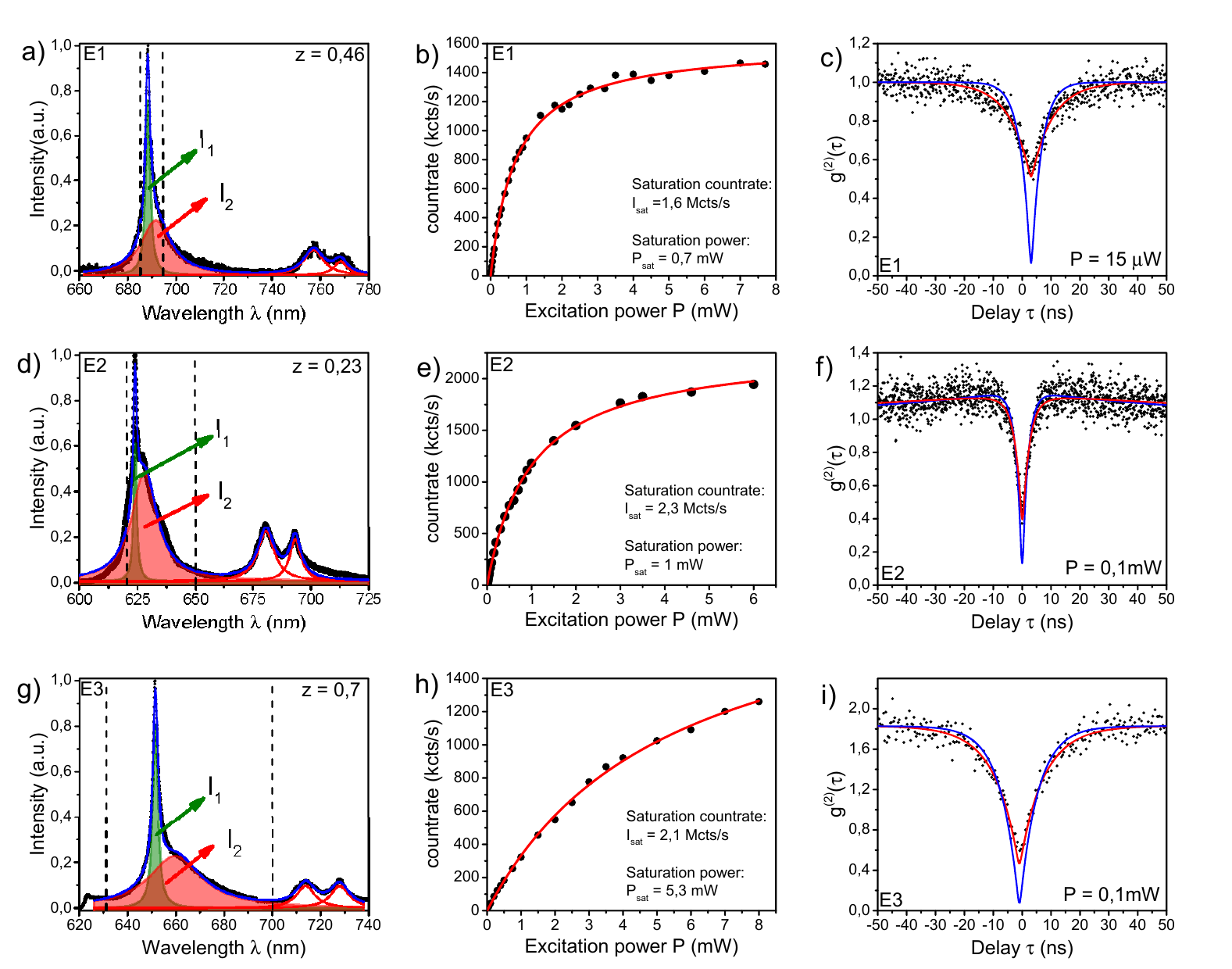}
 \caption{a,d,g) Typical spectra of three defects (E1,E2,E3) in
hBN consisting of four Lorentzian lines; b,e,f) Saturation
measurements on the defects from a),d) and g) with no
significant background contribution; c,f,i) g$^{(2)}$-intensity
correlation measurements on the defects shown in a),d) and g).
Photons are collected from the spectral regions enclosed in dashed lines in a),d) and g),
respectively. See main text for details.}
	\label{fig:g}
\end{figure}

\begin{figure}[ht]
	\centering
		\includegraphics[width=0.5\textwidth]{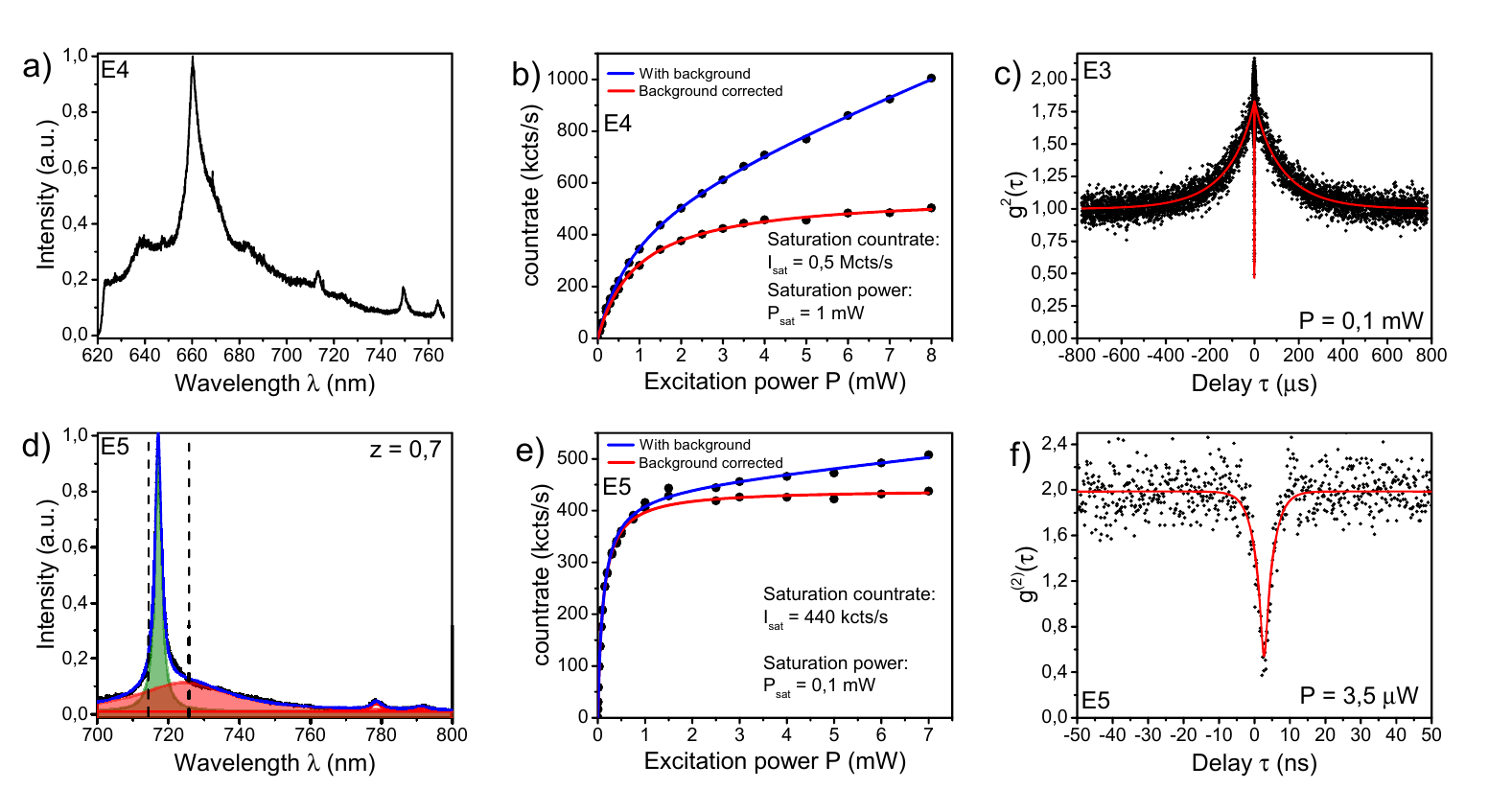}
 \caption{a) Spectrum of an emitter in hBN (E4) with a clear
background contribution; b) Saturation measurement on the
emitter in a). The background contribution is visible as
prominent linear increase in the emission rate at increasing
excitation powers; c) g$^{(2)}$-function on an hBN emitter (E3)
showing typical bunching timescales of several hundreds of
microseconds; d,e,f) Spectrum, Saturation measurement and
g$^{(2)}$-function of an emitter (E5) with a clean spectrum.
Background contribution becomes relevant at about
20xP$_{\text{sat}}$. Still, g$^{(2)}$(0) is strongly limited
even at almost vanishing excitation powers.}
	\label{fig:back}
\end{figure}

Fig.\ref{fig:g}a,d,g) show typical spectra of point emitters
inside the flakes.
Although they differ in their central wavelengths, their
spectral shapes are very similar. The spectra are fit with four
Lorentzian lines which we will discuss later in closer detail.
Saturation measurements in Fig.\ref{fig:g}b,e,h) show typical
saturation count rates ($\approx$ 1-2\,Mcts/s) and saturation
powers ($\approx$ 1\,mW) of these emitters, in good agreement to
previous reports \cite{Tran2015,Tran2016a,Tran2016b}. The red
lines are fits according to
\begin{equation}
I(P) = \frac{I_{\text{sat}}\cdot P}{P_{\text{sat}}+P} + C_{\text{back}}\cdot P.
\end{equation}
Here, $I_{\text{sat}}$ and $P_{\text{sat}}$ are the saturation
count rates and saturation powers of the emitters, whereas
$C_{\text{back}}$ describes a potential contribution due to
linear background emission stemming from the host material.
Note, that this contribution is negligible in the presented
data. This is in accordance with the very clean spectra
presented in Fig.\ref{fig:g}a,d,g), where also no significant
background contribution is visible. Contrary to these findings,
Fig.\ref{fig:back}a) shows a spectrum, which clearly contains
additional background emission. In approximately one out of
fifty flakes background-free emission can be found. This
background emission is also visible as a prominent linear
increase in a corresponding saturation measurement in
Fig.\ref{fig:back}b). Note, that the saturation measurements are
always taken including all four lines. As a last step, we
perform g$^{(2)}$-photon correlation measurements
(Fig.\ref{fig:g}c,f,i) to get information about the photon
statistics. Even though we did not observe any background
emission in all previous measurements, we further reduce the
spectral window from which we collect photons for the
g$^{(2)}$-measurements to the region of the ZPL (regions
enclosed by dashed lines in the corresponding spectra) and take
the measurements at excitation powers far lower than the
emitters' saturation powers. It strikes the eye, that despite
vanishing background fluorescence, the g$^{(2)}$-functions do
not vanish at all at zero time delay as one would expect for an
ideal single photon source. Fig.\ref{fig:back}d,e,f) further
shows an example, where background emission from the host
material is present but becomes relevant only at about
20xP$_{\text{sat}}$. Nevertheless, for almost vanishing
excitation power (P=3,5\,$\mu$W), the value of g$^{(2)}$(0) is
still much larger than zero. As we show below also the timing
jitter of the photon detectors does not explain the deviation
from ideal single photon statistics as the emitter fluorescence
lifetime is larger than the jitter. Instead, we have to assume
that the asymmetric shape of the ZPL is due to the presence of
two independent emission lines. \\
In the following we develop a model model for the photon
correlation functions that, besides background emission and the
timing jitter of the photon detector, accounts for the presence
of a second emitter and prove that this model fully reproduces
the measurements. We start with the well-known
g$^{(2)}$-function for a three level system:
\begin{equation}
g_i^{(2)}(\tau) = 1-(1+a)\cdot e^{-\frac{|\tau|}{\tau_1}}+a\cdot e^{-\frac{|\tau|}{\tau_2}}
\label{g2-dreiniveau}
\end{equation}
We now, step by step, include all experimental parameters, that
influence the shape of the g$^{(2)}$-function: Although
negligible in the presented data (but not in general), we start
with uncorrelated background emission, that can be extracted
from saturation measurements. Including this into the model, the
g$^{(2)}$-function reads \cite{Brouri2000}:
\begin{equation}
g_p^{(2)}(\tau) =
\frac{1}{p^2}\cdot\left[g_i^{(2)}(\tau)-(1-p^2)\right]
\label{g2pf}
\end{equation}
Here, $p$ is the fraction of measured photons stemming from the
emitter compared to the measured total count rate. Note, that
one should also consider dark counts of the detector in the
description. In our case, these dark counts ($\approx$ 100-200
cts/s) are negligible compared to the signal from the emitters.
Second, we include the timing jitter $\sigma$ of the counting
electronics. This jitter is an uncertainty in the time between
the arrival and the detection of a single photon and has been
measured via ultra-fast laser pulses ($\sigma \approx 490$\,ps).
It is included via the convolution of equation \ref{g2pf} with
the Gaussian-shape of the instrument response function IRF(t).
\begin{equation}
g_{p,j}^{(2)}(\tau) = \text{IRF}(\tau)*g_p^{(2)}(\tau) =
\int_{-\infty}^{\infty}\text{IRF}(\tau)\cdot g_p^{(2)}(\tau-t) dt
\label{eq:pj}
\end{equation}
Equation \ref{eq:pj} is the final description for the case that
we collect emission from exactly one single emitter. The blue
solid lines in Fig.\ref{fig:g}c,f,i) are fits to the data
according to this model. It strikes the eye that this function
is not able to reproduce the data. In particular, the model
demands a much lower value for g$^{(2)}$(0) than it is provided
by the data. We want to stress that we also can reproduce the
data by taking the signal to background ratio $p$ as a fit
parameter. This, however, strongly contradicts our findings of
vanishing background in the spectrum and the saturation
measurement.\\
Therefore, as a last step, we also take into account the
influence of a second emitter in the detection focal volume.

Let $I_{\text{tot}}=I_1+I_2$ be the total detected emission with
$I_1=z\cdot I_{\text{tot}}$ and $I_2=(1-z)\cdot I_{\text{tot}}$
being the relative fractions of the emission of emitter 1 and
emitter 2 respectively. This leads to
\begin{align*}
g^{(2)}(\tau)=\frac{\langle
I_{\text{tot}}(t)I_{\text{tot}}(t+\tau)\rangle}{\langle
I_{\text{tot}}(t)\rangle^2}\nonumber\\
= z^2\cdot g_1^2(\tau)+(1-z)^2\cdot g_2^2(\tau) + \\
\underbrace{\frac{\langle I_2(t)I_1(t+\tau)\rangle}{\langle
I_{\text{ges}}(t)\rangle^2}+\frac{\langle I_1(t)I_2(t+
\tau)\rangle}{\langle
I_{\text{ges}}(t)\rangle^2}}_{g^2_{\text{mix}}}\nonumber.
\end{align*}
In order to reduce the number of fit parameters, we assume
g$_1^2(\tau) =$ g$_2^2(\tau)$. Because of the independence of
$I_1$ and $I_2$, the two mixing terms will be constant for all
$\tau$ and by making the assumption, that g$_1^2(0) =$
g$_2^2(0)$=$0$, we find g$^2_{\text{mix}}=2z(1-z)$. We
eventually arrive at
\begin{equation}
g^{(2)}(\tau) = (1-2z(1-z))g_{p,j}^{(2)}(\tau) + 2z(1-z).
\label{eq:Full}
\end{equation}
In contrast to reports in literature, where the asymmetry of the
ZPL is attributed to phonon interaction \cite{Tran2015}, we here
fully reproduce the lineshape by fitting two Lorentzian lines
representing two independent electronic transitions. By
calculating the areas under the individual Lorentzians, we get
information about the relative oscillator strengths of both
emitters, corresponding to the parameter $z$ in equation
\ref{eq:Full} (numbers also given in the spectra in
Fig.\ref{fig:g} and Fig.\ref{fig:back}). By taking into account
the double emission spectrum within the model for the
g$^{(2)}$-function we are able to perfectly describe the
measured photon correlation data (solid red lines in
Fig.\ref{fig:g}c,f,i and Fig.\ref{fig:back}c,f).

\newpage

\begin{figure}[ht]
	\centering
		\includegraphics[width=0.5\textwidth]{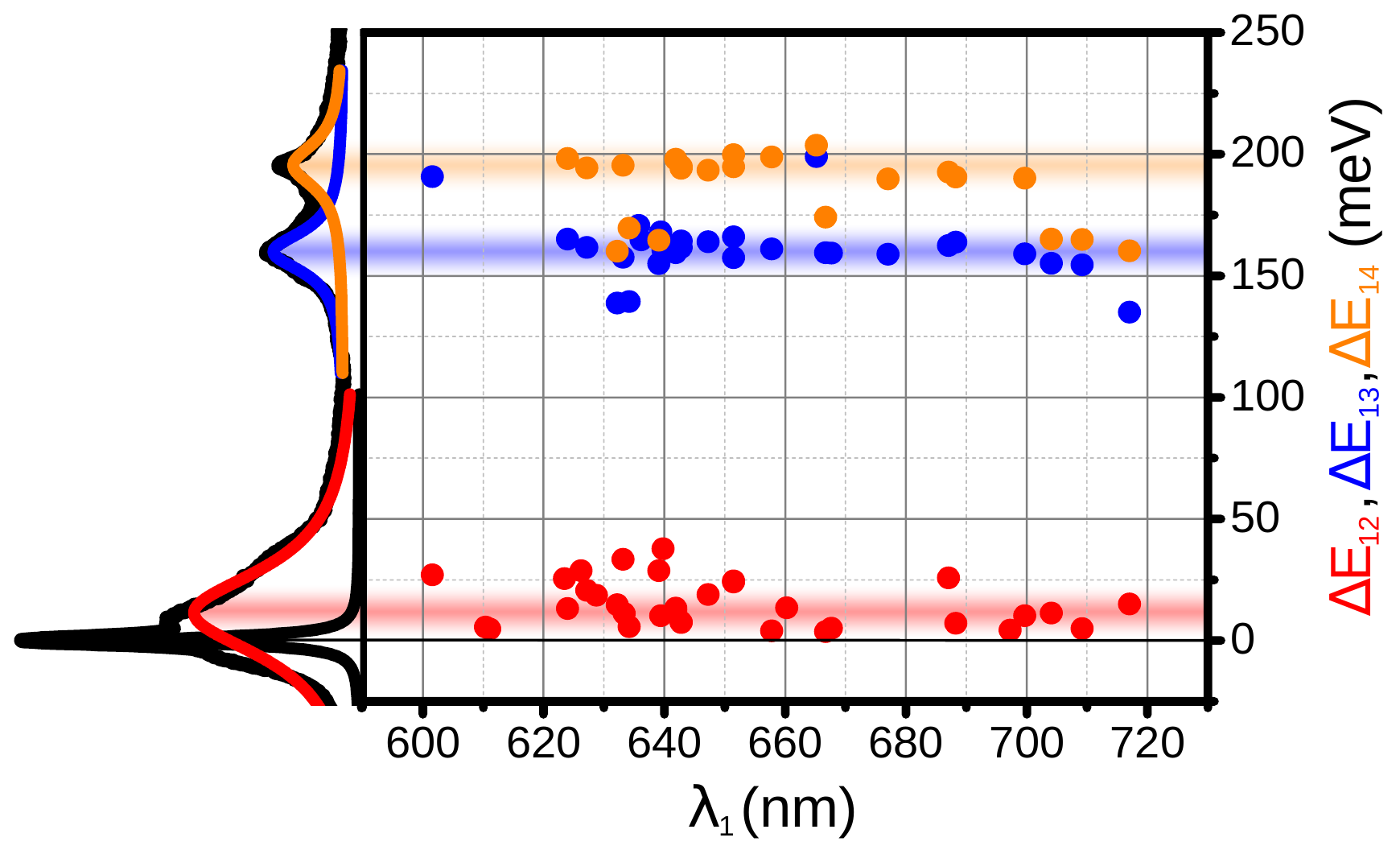}
 \caption{Spectral line position data extracted from 30
emitters that show comparable spectral fingerprints as explained
in the main text. Due to strain in the material, the central
wavelengths of peak 1 range from 600\,nm - 720\,nm. However,
independent of the wavelengths, the energy distance between the
lines roughly remain constant. $\Delta$E$_{12}$ = 12(10)\,meV,
$\Delta$E$_{13}$ = 158(17)\,meV, $\Delta$E$_{14}$ =
187(15)\,meV, $\Delta$E$_{24}$ = 174(15)\,meV. The
semi-transparent, horizontal lines are a guide to the eye.}
	\label{fig:spec}
\end{figure}

Interestingly, our photon correlation measurements correspond
perfectly to reports in literature in terms of bunching dynamics
and dips in the g$^{(2)}$-function at zero time delay
\cite{Tran2015,Tran2016a,Tran2016b}. Non-vanishing values of
g$^{(2)}$(0) in these reports were always attributed to residual
background fluorescence which, however, is not further defined
or shown. To our knowledge, the full set of information needed
to accurately describe the situation has never been reported
\cite{Tran2015,Tran2016b,Mendelson2018,Choi2016}. Furthermore,
we want to point out that most of the emitters measured in this
work show very strong bunching on a timescale of several
hundreds of microseconds up to milliseconds as it has been shown
in previous work (see for example Fig.\ref{fig:back}c)
\cite{Tran2016a,Tran2016b}. Therefore, a proper normalization of
the g$^{(2)}$-function to the constant number of events for long
time delays $\tau$ or to the recorded photon count rates is
imperative. The absence of satisfactorily explanations in
literature and the excellent agreement of measured photon
correlation functions with the double defect model suggest that
most probably the majority of \textit{single} emitters in
literature are indeed double defects.\\
We now turn in closer detail to the emitters optical spectra
providing further evidence for our model. Fig.\ref{fig:spec}
shows normalized emission spectra of a collection of emitters in
the multi-layer flakes under investigation. The central
wavelength $\lambda_1$ of the highest energy line (line 1)
ranges from 600-720\,nm. This wavelength range is limited by the
spectral filter window in which we collect fluorescence. On the
y-axis the energy separations between all lines are shown, where
the energy of line 1 (black) is always set to zero. It strikes
the eye, that the energy distances between the lines remain
approximately constant independent of the central wavelength of
line 1 in the spectrum. In literature, the spectrum is described
as an asymmetric zero phonon line with a red shifted (165\,meV)
phonon side band the energy of which belongs to a well-known
phonon mode in hBN \cite{Geick1966,Reich2005,Nemanich1981}. We
here first state that there are actually two ZPLs (line 1, black
and line 2, red) with two phonon side bands (line 3, blue and
line 4, orange). Averaged over all observed emitters with this
particular spectral fingerprint, the energy difference between line 1
(black) and line 3 (blue) amounts to
$\Delta$E$_{13}$=158(17)\,meV, whereas the distance between line
2 (red) and line 4 (orange) yields
$\Delta$E$_{24}$=174(15)\,meV. Within the error bars, both
values match the phonon mode at 165\,meV. Staying in our picture
of line 1 and 2 being electronic transitions, we thus attribute
the lines 3 and 4 to be their respective phonon side bands.

\begin{figure}[h!]
	\centering
				\includegraphics[width=0.5\textwidth]{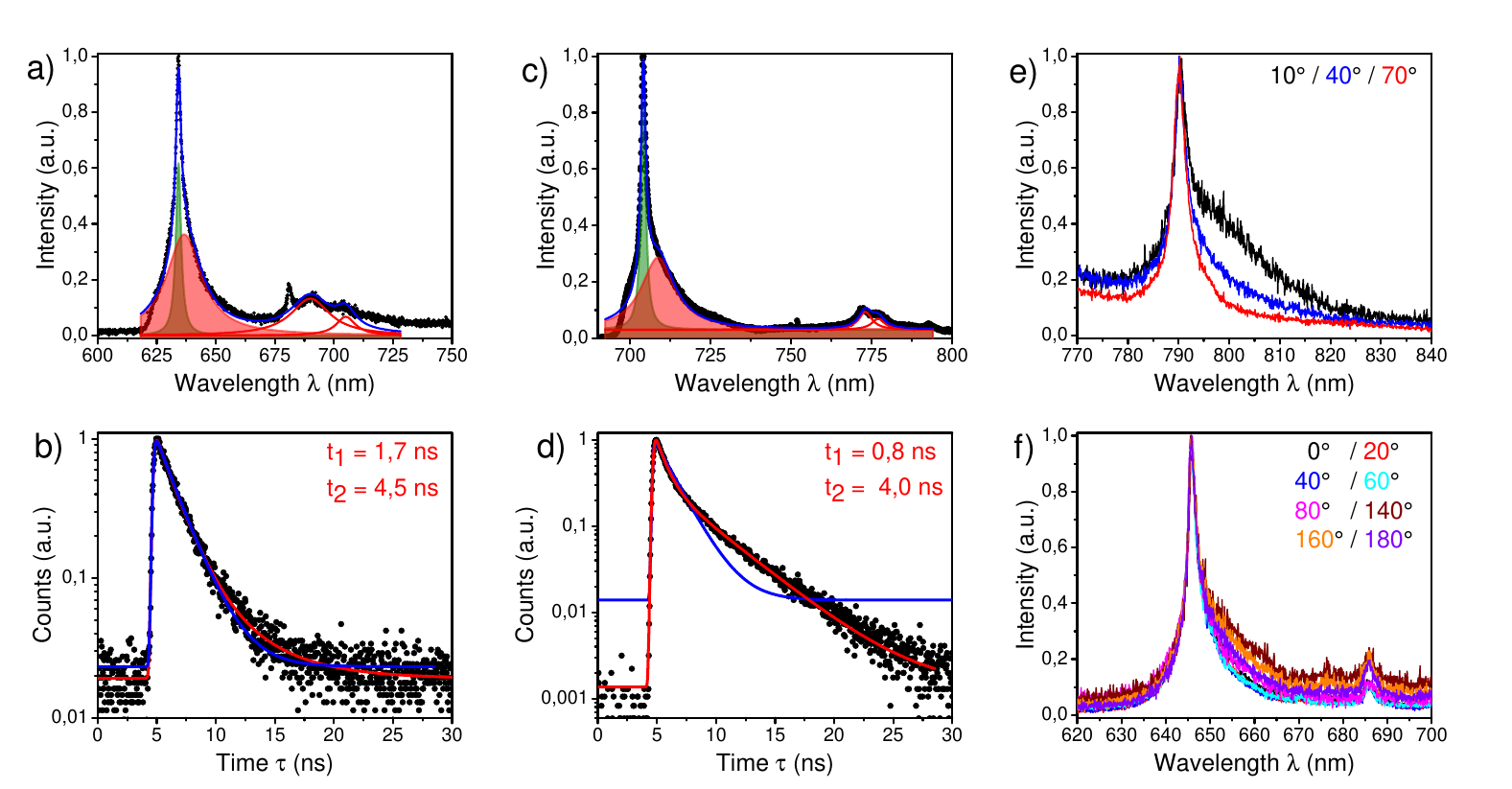}
 \caption{a,c) Spectra of two typical emitter in hBN. The
solid, colored lines are Lorentzian fits to the data. b,d)
Lifetime measurements on the emitters shown in a) and c). The
solid lines are fits according to a mono-exponential (blue) and
bi-exponential (red) decay including the instrument response
function of the setup. Both measurements follow a bi-exponential
decay with time constants $t_1$=0.82\,ns, $t_2$=4.0\,ns and
$t_1$=1.7\,ns, $t_2$=4.5\,ns. e,f) Polarization dependent
optical spectra of two typical hBN emitter in emission. The line
shape strongly depends on the angle of an polarizer in the
detection path.}
	\label{fig:LP}
\end{figure}

Next, we perform TCSPC-measurements to gain further information
about the lifetimes of the excited states of the investigated
emitters. Two electronic transitions with potentially differing
lifetimes should be visible as a bi-exponential decay. For the
TCSPC-measurements we use a white light laser filtered to
532\,nm, with a pulse duration of 200\,ps and a pulse repetition
rate of 10\,MHz. Measurements on two typical hBN emitters
(Fig.\ref{fig:LP}a,c) are shown in Fig.\ref{fig:LP}b,d) (black
dots). The solid lines are fits according to
\begin{equation}
L(t) =
y_0+\left(1-\text{erf}\left(-\frac{t-t_0}{\sigma}\right)\right)\cdot\sum_{i=1}^n
A_i\cdot \displaystyle{\text{e}^{-\frac{t-t_0}{t_i}}}
\end{equation}
with one (blue, $n$=1) and two (red, $n$=2) time constants. As
for the g$^{(2)}$-functions the instrument response function
IRF(t) with a timing jitter of $\sigma$=490\,ps is included into
the fit function via convoluting the Gaussian IRF(t) with the
exponential decay of the electronic transition. In both
measurements, the data points clearly follow a bi-exponential
decay. The observed time constants ($t_1$=0.82\,ns,
$t_2$=4.0\,ns and $t_1$=1.7\,ns, $t_2$=4.5\,ns) correspond to
the range of typical lifetimes observed for this type of
emitters \cite{Tran2015,Tran2016a,Tran2016b}. In literature,
however, usually just a single exponential decay is used to fit
the data in a regime between 2-10\,ns and the full information
about the timing resolution of the setup is not considered.
The presence of two time constants of the same order of
magnitude further indicates the existence of two excited states
in the defect and corresponds perfectly with the assumption of
two independent emitters in the same defect.\\
As a last step we now turn to the polarization of the defect
emission. Linear excitation dipoles with visibilities between
20\,\% and 80\,\% have been reported whereas the emission dipole
are supposed to show close to unity visibility and are linearly
polarized \cite{Tran2015,Exarhos2017}. There is, however, a
difference in the relative orientations of the excitation and
emission dipoles between 30$^{\circ}$ to 90$^{\circ}$
\cite{Exarhos2017,Choi2016}. However, to our knowledge,
polarization dependent spectra have not been investigated in
literature yet. Fig. \ref{fig:LP}e,f) show normalized emission
spectra of two different hBN emitters with ZPL (line 1) at
around 790\,nm and 650\,nm where different curves correspond to
different settings of the polarization analyzer in the detection
path. One can clearly see that dependent on the angle of a
linear polarizer in the detection path, the line shape of the
dominant line in the spectrum strongly varies. This indicates
that here the two dipoles contributing to the ZPL have different
relative polarizations. Note, that we also can find spectra in
which the line shape does not change significantly upon changing
the detection angle of the polarization analyzer.
\section{Conclusions}
In summary, we presented new insights into the nature of
non-classical light emission from defects in multi-layer flakes
of hexagonal boron nitride. Via careful evaluation of
g$^{(2)}$-photon correlation measurements, TCSPC-measurements
and polarization dependent emission spectra we gather strong
evidence that, in contrast to previous reports, these atomic
defects are no single emitter systems but are comprised of two
independent emitting systems here coined as "`double defect"'.
We draw this conclusion via collecting all necessary information
to describe the photon statistics through independent
measurements of the background contribution, the timing jitter
of the counting electronic and the spectra of the emitters.
Interestingly, our photon correlation measurements correspond
perfectly to reports in literature in terms of bunching dynamics
and dips in the g$^{(2)}$-function at zero time delay. \\
Our assumptions are corroborated by the decomposition of the
asymmetric ZPL into two Lorentzian lines, both describing one
individual electronic transition. Based on the existence of a
characteristic phonon mode of hBN at 165\,meV, we were able to
assign the two dominant lines in the phonon side band
to each of the electronic transitions. Eventually, the presence
of a bi-exponential decay in TCSPC measurements and polarization
dependent emission spectra further support our model. We want to
point out that our measured photon correlation functions
perfectly correspond to the ones previously reported in
literature. Based on these results we have to assume that many of the reported single photon emitters
consist of "double defects"' as described in this
publication.
\section{Acknowledgement}
The authors want to thank Johannes G\"orlitz, Benjamin Kambs,
Dennis Herrmann, Igor Aharonovich, Dirk Englund, Lee Bassett and
Adam Gali for helpful discussions. This work was partially funded by the
European Union 7th Framework Program under Grant Agreement No. 61807 (WASPS).

\bibliographystyle{}

\end{document}